\documentclass{article}
\usepackage{spconf,amsmath,graphicx,hyperref,amssymb}
\usepackage{enumitem}
\usepackage{booktabs}
\usepackage{multirow}
\usepackage{siunitx}
\usepackage{makecell} 

\usepackage{subcaption} 
\title{Exploring Fine-Tuning of Large Audio Language Models for Spoken Language Understanding under Limited Speech Data}
\name{Youngwon Choi$^{1}$, 
Jaeyoon Jung$^{1,2}$, 
Hyeonyu Kim$^{1}$, 
Huu-Kim Nguyen$^{1}$\sthanks{Now with Atmanity Inc., USA.},
Hwayeon Kim$^{1}$}
\address{
  $^{1}$MAUM AI Inc., Republic of Korea \\
  $^{2}$Soongsil University, Republic of Korea 
}

\begin{document}
%
\maketitle
\begin{abstract}
Large Audio Language Models (LALMs) have emerged as powerful tools for speech-related tasks but remain underexplored for fine-tuning, especially with limited speech data. 
To bridge this gap, we systematically examine how different fine-tuning schemes including text-only, direct mixing, and curriculum learning affect spoken language understanding (SLU), focusing on scenarios where text–label pairs are abundant while paired speech–label data are limited.
Results show that LALMs already achieve competitive performance with text-only fine-tuning, highlighting their strong generalization ability. 
Adding even small amounts of speech data (2–5\%) yields substantial further gains, with curriculum learning particularly effective under scarce data. In cross-lingual SLU, combining source-language speech data with target-language text and minimal target-language speech data enables effective adaptation. 
Overall, this study provides practical insights into the LALM fine-tuning under realistic data constraints.

\end{abstract}
\section{Introduction}
\label{sec:intro}

Large Audio Language Models (LALMs) such as AudioPaLM \cite{rubenstein2023audiopalm}, SALMONN~\cite{TangEtAl2024SALMONN}, and Qwen-Audio~\cite{Qwen-Audio} have recently emerged. 
By integrating pretrained audio encoders with large language models, LALMs unify acoustic and linguistic representations in an end-to-end manner, avoiding the limitations of ASR-to-LLM cascades in speech-related tasks, such as transcription error propagation~\cite{fathullah2024audiochatllama, shang2024end} and the loss of paralinguistic information~\cite{sarim2025direct, wu2024just}.

While most studies of LALMs have focused on large-scale pre-training and zero-shot evaluation, research on fine-tuning is still at an early stage, in contrast to the text LLM literature where fine-tuning has become the standard paradigm for downstream adaptation~\cite{shengyu2023instruction}. Recent works have explored this direction, including domain-specific applications~\cite{florea2025exploring, bn2025fine} and cross-task spoken language understanding (SLU) adaptation through in-context learning~\cite{agrawal2025spoken}. Omni-R1~\cite{rouditchenko2025omni} further showed that even text-only fine-tuning can enhance audio QA performance, suggesting that curriculum training with text-only and audio-text data may be a promising direction for future work.
Taken together, these efforts highlight the emerging interest in off-the-shelf LALM fine-tuning, yet how such approaches can be systematically applied to downstream speech processing tasks under realistic speech resource constraints remains an open question.

In this work, we adopt SLU as a benchmark task to evaluate the practical effectiveness of LALM fine-tuning. 
SLU is a key task in speech processing and a core component of conversational agents and voice assistants~\cite{tur2011spoken,bastianelli2020slurp}, while also representing a low-resource setting where paired speech–label data are expensive to collect and annotate~\cite{huang2020leveraging,qin2021survey}.
Prior studies have proposed several approaches to address this limitation; one such approach is transcription-based training~\cite{thomas2022towards}, leveraging transcripts that are easier to obtain than speech data and can partially compensate for scarcity.

We argue that LALMs are particularly well positioned for this setting, as their strong language modeling capability and pre-trained audio–text alignment allow them to incorporate limited speech data while exploiting abundant text resources. 
Based on this motivation, we summarize our contributions as follows:
(i) We demonstrate that LALMs can achieve strong SLU performance with transcript-based training combined with limited speech data, providing a cost-efficient path for adaptation;
(ii) We establish a systematic benchmark of fine-tuning schemes for LALMs across different speech data budgets and show that curriculum learning is more effective with limited data, while direct mixing suffices once more data is available;
(iii) We further show that in low-resource languages, LALMs can be effectively adapted cross-lingually by combining source-language speech data with target-language text and minimal target-language speech data.



\begin{figure*}[t]
  \centering
  \begin{minipage}[b]{0.55\textwidth}
    \centering
    \includegraphics[width=\linewidth]{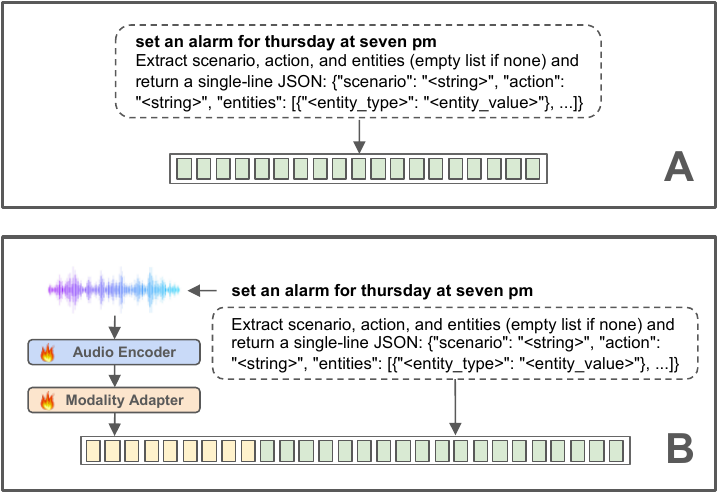}\par\medskip
    \vspace{-0.5em}
    {\small\textbf{(a)} End-to-end SLU objective and prompting scheme.}
    \vspace{-0.3em}
  \end{minipage}\hfill
  \begin{minipage}[b]{0.45\textwidth}
    \centering
    \includegraphics[width=0.8\linewidth]{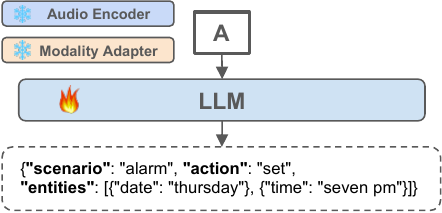}\par\medskip
    {\small\textbf{(b)} Text-only}
    \vspace{0.9em}
    \includegraphics[width=0.8\linewidth]{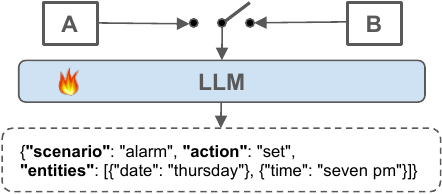}\par\medskip
    \vspace{-0.9em}
    {\small\textbf{(c)} Direct mixing}
    \vspace{-0.3em}
  \end{minipage}

  \caption{Comparison of fine-tuning schemes for SLU. (a) shows the unified task/prompt format. (b) illustrates the text-only, and (c) the direct mixing. Curriculum learning applies (b) in early epochs and (c) in the final epoch.}
  \vspace{-0.7em}
  \label{fig:schemes}
\end{figure*}

\section{Fine-tuning LALMs for SLU}
\label{sec:format}

\subsection{Task Definition and Evaluation}

Following prior work~\cite{huang2023leveraging}, we cast SLU as a joint task of intent classification and slot filling, where each intent consists of a scenario and an action. In our setup, the model directly maps speech to semantic labels without relying on intermediate ASR transcripts, enabling an end-to-end evaluation of the SLU performance of LALMs. For training targets, we represent the semantics as a structured sequence that combines the intent and its slot–filler pairs, as in Fig.~\ref{fig:schemes}(b) and (c).

To assess the effectiveness of fine-tuning LALMs for SLU, we adopt three standard metrics: (i) \textit{intent accuracy}, the proportion of utterances with correctly predicted intents; (ii) \textit{entity F1}, the micro-averaged F1 over entity spans; and (iii) \textit{SLU-F1}~\cite{bastianelli2020slurp}, an utterance-level score jointly assessing intent and entity prediction.

\subsection{Adaptation of LALMs under Scarce Resources}
\label{ssec:subhead}

We analyze how LALMs can be fine-tuned for SLU under limited speech data. To this end, we compare three fine-tuning schemes, which differ in how speech data is incorporated during training. We experiment in a \emph{monolingual SLU} setting, where both training and evaluation are conducted in a single language.

\begin{itemize}[leftmargin=*, itemsep=0pt, topsep=0pt]
\item 
    \textbf{Text-only}: training is performed using ground-truth transcripts with semantic labels, while the audio encoder and modality adapter layers are kept frozen and only the LLM weights are updated. 
\item 
    \textbf{Direct mixing}: training combines all transcripts with a fixed proportion of speech–label pairs (ranging from 2\% to 100\% of the corpus), which are randomly sampled and mixed in the training batches; the entire LALM is updated in this setting.
\item 
    \textbf{Curriculum Learning}: training relies on the same data as direct mixing, but applies it in two phases. Training begins with transcripts only, and speech data is introduced only in the final epoch, following the same manner as direct mixing.
    To ensure fairness, both direct mixing and curriculum learning expose the model to the same total amount of speech data, $N_p = pN$, where $N$ is the number of speech samples in the corpus and $p$ is the proportion selected. With $E$ training epochs, the per-epoch allocation is
    \[
    a_e^{\text{direct}} = N_p, \quad 
    a_e^{\text{curr}} = 
    \begin{cases}
        0, & 1 \leq e < E, \\
        N_p \times E, & e = E,
    \end{cases}
    \]
    where $a_e$ denotes the number of speech–label pairs used at epoch $e$.
\end{itemize}

\begin{table}[t]
\centering
\caption{Corpora used in our experiments.
$\ast$ For the \emph{cross-lingual SLU} setting, we use 115 paired transcript–speech samples for target languages.}
\label{tab:datasets}
\resizebox{0.98\linewidth}{!}{%
\begin{tabular}{llcccccc}
\toprule
Corpus & Lang & \multicolumn{2}{c}{Train} & \multicolumn{2}{c}{Dev} & \multicolumn{2}{c}{Test} \\
\cmidrule(lr){3-4}\cmidrule(lr){5-6}\cmidrule(lr){7-8}
 & & \#text & \#speech & \#text & \#speech & \#text & \#speech \\
\midrule
SLURP & EN & 11,514 & 50,628 & 2,033 & 8,690 & 2,974 & 13,078 \\
ITALIC & IT & 11,514 & 11,514 & 2,033 & 2,033 & 2,974 & 2,974 \\
\multirow{2}{*}{S-MASSIVE} 
 & FR       & 11,514 & 11,514 & 2,033 & 2,033 & 2,974 & 2,974 \\
 & 11 langs & 115$\ast$ & 115$\ast$ & 2,033 & 2,033 & 2,974 & 2,974 \\
\bottomrule
\end{tabular}%
}
\end{table}

\setlength{\tabcolsep}{3pt} 
\begin{table*}[t]
\centering
\caption{Monolingual results on 3 datasets. We report intent accuracy, entity F1, and SLU-F1 with 95\% confidence intervals.}
\label{tab:monolingual}
\resizebox{\linewidth}{!}{
\begin{tabular}{c l | c c c | c c c | c c c}
\toprule
\multirow{2}{*}{Speech} & \multirow{2}{*}{Scheme} 
& \multicolumn{3}{c|}{SLURP} 
& \multicolumn{3}{c|}{ITALIC} 
& \multicolumn{3}{c}{Speech-MASSIVE FR} \\
& & Intent Accuracy & Entity~F1 & SLU-F1 
  & Intent Accuracy & Entity~F1 & SLU-F1
  & Intent Accuracy & Entity~F1 & SLU-F1 \\
\midrule
& \emph{Oracle} 
& 0.9129 $\pm$ 0.0044 & 0.8359 $\pm$ 0.0057 & 0.8593 $\pm$ 0.0047
& 0.8830 $\pm$ 0.0035 & 0.7658 $\pm$ 0.0041 & 0.8095 $\pm$ 0.0042 
& 0.8854 $\pm$ 0.0034 & 0.7541 $\pm$ 0.0054 & 0.7985 $\pm$ 0.0039 \\
\midrule
0\% & Text & 0.8360 $\pm$ 0.0018 & 0.6406 $\pm$ 0.0026 & 0.7207 $\pm$ 0.0017
           & 0.7834 $\pm$ 0.0036 & 0.5661 $\pm$ 0.0063 & 0.6755 $\pm$ 0.0028
           & 0.8017 $\pm$ 0.0033 & 0.5130 $\pm$ 0.0130 & 0.6535 $\pm$ 0.0071 \\
\multirow{2}{*}{2\%} & Direct & 0.8345 $\pm$ 0.0082 & 0.6354 $\pm$ 0.0038 & 0.7167 $\pm$ 0.0038
             & 0.8048 $\pm$ 0.0147 & 0.5644 $\pm$ 0.0152 & 0.6773 $\pm$ 0.0098
             & 0.8132 $\pm$ 0.0078 & 0.5349 $\pm$ 0.0113 & 0.6740 $\pm$ 0.0098 \\
     & Curr. & \textbf{0.8574 $\pm$ 0.0033}$^{\dagger}$ & \textbf{0.6577 $\pm$ 0.0024}$^{\dagger}$ & \textbf{0.7335 $\pm$ 0.0011}$^{\dagger}$
             & \textbf{0.8272 $\pm$ 0.0029}$^{\dagger}$ & \textbf{0.6074 $\pm$ 0.0065}$^{\dagger}$ & \textbf{0.7088 $\pm$ 0.0044}$^{\dagger}$
             & \textbf{0.8287 $\pm$ 0.0077} & \textbf{0.5590 $\pm$ 0.0063}$^{\dagger}$ & \textbf{0.6919 $\pm$ 0.0048}$^{\dagger}$ \\
\multirow{2}{*}{5\%} & Direct & 0.8558 $\pm$ 0.0050 & 0.6617 $\pm$ 0.0054 & 0.7373 $\pm$ 0.0042
             & 0.8376 $\pm$ 0.0048 & 0.6190 $\pm$ 0.0070 & 0.7165 $\pm$ 0.0037
             & 0.8376 $\pm$ 0.0118 & 0.5677 $\pm$ 0.0053 & 0.6969 $\pm$ 0.0037 \\
     & Curr. & \textbf{0.8642 $\pm$ 0.0016}$^{\dagger}$ & \textbf{0.6765 $\pm$ 0.0021}$^{\dagger}$ & \textbf{0.7475 $\pm$ 0.0025}$^{\dagger}$
             & \textbf{0.8412 $\pm$ 0.0032} & \textbf{0.6334 $\pm$ 0.0072}$^{\dagger}$ & \textbf{0.7271 $\pm$ 0.0041}$^{\dagger}$
             & \textbf{0.8423 $\pm$ 0.0023} & \textbf{0.5802 $\pm$ 0.0035}$^{\dagger}$ & \textbf{0.7048 $\pm$ 0.0044} \\
\multirow{2}{*}{10\%} & Direct & 0.8618 $\pm$ 0.0023 & 0.6740 $\pm$ 0.0031 & 0.7482 $\pm$ 0.0022
             & \textbf{0.8533 $\pm$ 0.0054} & 0.6387 $\pm$ 0.0060 & 0.7320 $\pm$ 0.0034
             & 0.8418 $\pm$ 0.0066 & 0.5805 $\pm$ 0.0102 & 0.7054 $\pm$ 0.0059 \\
     & Curr. & \textbf{0.8678 $\pm$ 0.0037} & \textbf{0.6807 $\pm$ 0.0015}$^{\dagger}$ & \textbf{0.7529 $\pm$ 0.0021}$^{\dagger}$
             & 0.8490 $\pm$ 0.0034 & \textbf{0.6492 $\pm$ 0.0061} & \textbf{0.7406 $\pm$ 0.0022}$^{\dagger}$
             & \textbf{0.8493 $\pm$ 0.0017} & \textbf{0.5994 $\pm$ 0.0072}$^{\dagger}$ & \textbf{0.7174 $\pm$ 0.0056}$^{\dagger}$ \\
\multirow{2}{*}{25\%} & Direct & 0.8689 $\pm$ 0.0028 & 0.6779 $\pm$ 0.0040 & 0.7515 $\pm$ 0.0043
             & 0.8618 $\pm$ 0.0071 & 0.6650 $\pm$ 0.0041 & 0.7518 $\pm$ 0.0021
             & 0.8619 $\pm$ 0.0053 & 0.6150 $\pm$ 0.0065 & 0.7278 $\pm$ 0.0053 \\
     & Curr. & \textbf{0.8743 $\pm$ 0.0026} & \textbf{0.6873 $\pm$ 0.0023}$^{\dagger}$ & \textbf{0.7580 $\pm$ 0.0015}$^{\dagger}$
             & \textbf{0.8622 $\pm$ 0.0017} & \textbf{0.6690 $\pm$ 0.0039} & \textbf{0.7529 $\pm$ 0.0040}
             & \textbf{0.8634 $\pm$ 0.0024} & \textbf{0.6176 $\pm$ 0.0050} & \textbf{0.7285 $\pm$ 0.0037} \\
\multirow{2}{*}{50\%} & Direct & \textbf{0.8779 $\pm$ 0.0027} & \textbf{0.6891 $\pm$ 0.0028} & \textbf{0.7618 $\pm$ 0.0012}
             & 0.8680 $\pm$ 0.0025 & 0.6827 $\pm$ 0.0072 & 0.7597 $\pm$ 0.0042
             & 0.8708 $\pm$ 0.0050 & 0.6271 $\pm$ 0.0051 & 0.7363 $\pm$ 0.0049 \\
     & Curr. & 0.8771 $\pm$ 0.0033 & 0.6890 $\pm$ 0.0023 & 0.7610 $\pm$ 0.0022
             & \textbf{0.8687 $\pm$ 0.0040} & \textbf{0.6850 $\pm$ 0.0043} & \textbf{0.7616 $\pm$ 0.0033}
             & 0.8682 $\pm$ 0.0034 & \textbf{0.6311 $\pm$ 0.0033} & \textbf{0.7381 $\pm$ 0.0024} \\
\multirow{2}{*}{100\%} & Direct & \textbf{0.8813 $\pm$ 0.0035} & \textbf{0.6959 $\pm$ 0.0030} & \textbf{0.7675 $\pm$ 0.0022}
              & \textbf{0.8767 $\pm$ 0.0016} & 0.7022 $\pm$ 0.0054 & 0.7737 $\pm$ 0.0063
              & \textbf{0.8739 $\pm$ 0.0040} & 0.6445 $\pm$ 0.0031 & 0.7486 $\pm$ 0.0025 \\
       & Curr.  & 0.8810 $\pm$ 0.0043 & 0.6932 $\pm$ 0.0030 & 0.7644 $\pm$ 0.0017
              & 0.8735 $\pm$ 0.0034 & \textbf{0.7056 $\pm$ 0.0049} & \textbf{0.7766 $\pm$ 0.0032}
              & 0.8718 $\pm$ 0.0036 & \textbf{0.6463 $\pm$ 0.0075} & \textbf{0.7492 $\pm$ 0.0067} \\
\bottomrule
\end{tabular}}
\vspace{0.2em}
\footnotesize $^{\dagger}$ Significant improvement over the other method (95\% CI non-overlapping).

\vspace{-6pt}
\end{table*}

We further investigate \emph{cross-lingual SLU} setting to evaluate how abundant source-language data can be leveraged to transfer across languages and support SLU in low-resource languages. In the zero-shot setting, models are trained on source-language data using either text-only or curriculum learning schemes. The models are then evaluated on spoken utterances from multiple target languages, with no exposure to any target-language data during training. In the few-shot setting, training is performed on both the source-language data and a small amount of target-language data, which may consist of transcripts alone, or a combination of transcripts and paired speech–label examples. This setting directly evaluates the ability of LALMs to achieve generalizable SLU under scarce target-language resources.

In all cases, we provide a short instructional prompt, as shown in Fig.~\ref{fig:schemes}(a), which specifies the expected output structure (scenario, action, entities) to guide SLU adaptation.

\section{Experiments}
\label{sec:pagestyle}

\subsection{Datasets}
\label{ssec:datasets}

We conduct experiments on three corpora, as summarized in Table~\ref{tab:datasets}. For English SLU, we use SLURP~\cite{bastianelli2020slurp}, restricting to the real subset to align with the limited-speech scenario. To assess generalization beyond English, we also include ITALIC~\cite{koudounas2023italic}, an Italian SLU dataset.
Finally, we employ Speech-MASSIVE~\cite{lee2024speech}, a multilingual SLU corpus where the French subset is used for both \emph{monolingual SLU} and as the source language for transfer to 11 target languages in \emph{cross-lingual SLU}.


\subsection{Training and Inference Details}
\label{ssec:details}

We fine-tune Qwen2-Audio-7B-Instruct~\cite{chu2024qwen2} using the AdamW optimizer with bfloat16 precision. Training is conducted on eight H100 80GB GPUs with a per-device batch size of 2 and gradient accumulation of 8. All experiments run for three epochs, which we found sufficient for convergence in preliminary validation.

For text-only and direct mixing fine-tuning, we apply a cosine learning rate schedule, with a peak learning rate of $5.0\times10^{-6}$ and a warmup ratio of 0.04, across three epochs. For curriculum learning, the first two epochs are text-only using the same scheduler configuration, followed by a final epoch with both text and audio, using a reduced peak learning rate of $3.0\times10^{-6}$ and a warmup ratio of 0.02 to stabilize adaptation.

We evaluate six levels of speech data usage—2\%, 5\%, 10\%, 25\%, 50\%, and 100\%—where each larger level subsumes the smaller ones to ensure comparability across resource levels. For inference, we use beam search with three beams.

In the Monolingual setting, we report results averaged over five random seeds and compute 95\% confidence intervals using the t-distribution to assess that observed gaps between fine-tuning schemes are reliable.

\subsection{Results on Monolingual SLU}
\label{ssec:results_mono}

Table~\ref{tab:monolingual} summarizes \emph{monolingual SLU} results on SLURP, ITALIC, and Speech-MASSIVE FR.
It also includes an \emph{oracle} setting, obtained with the text-only scheme and gold transcripts at inference time, which provides a substantial upper bound for current E2E SLU models. 
The text-only baseline already achieves strong performance, reaching about 85–95\% of the peak SLU-F1 scores across corpora (about 94\% on SLURP and 87\% on ITALIC and Speech-MASSIVE FR).
Nevertheless, adding as little as 2–5\% paired speech data yields large gains, quickly raising SLU-F1 to 97\% of the peak on SLURP and 94\% on others, showing that even a small amount of speech provides valuable grounding.

At low levels of speech data (2–10\%), curriculum training broadly outperforms direct mixing.
In SLU-F1, the 95\% confidence intervals of the two methods are non-overlapping across datasets, except for a negligible overlap in Speech-MASSIVE FR at 5\%.
This suggests that stabilizing the semantic task with text-only training before introducing speech data helps the model integrate acoustic information more effectively and mitigates overfitting under scarce resources, making curriculum learning a more reliable strategy in the low-resource regime.

When the amount of speech data increases to 25\% or more, the gap between the two schemes narrows considerably; only SLURP at 25\% shows a clear difference.
With abundant speech resources, the two schemes converge, and direct mixing proves to be a stable and sufficient approach without curriculum learning.

\begin{figure}[t]
    \centering
    \includegraphics[width=0.94\linewidth]{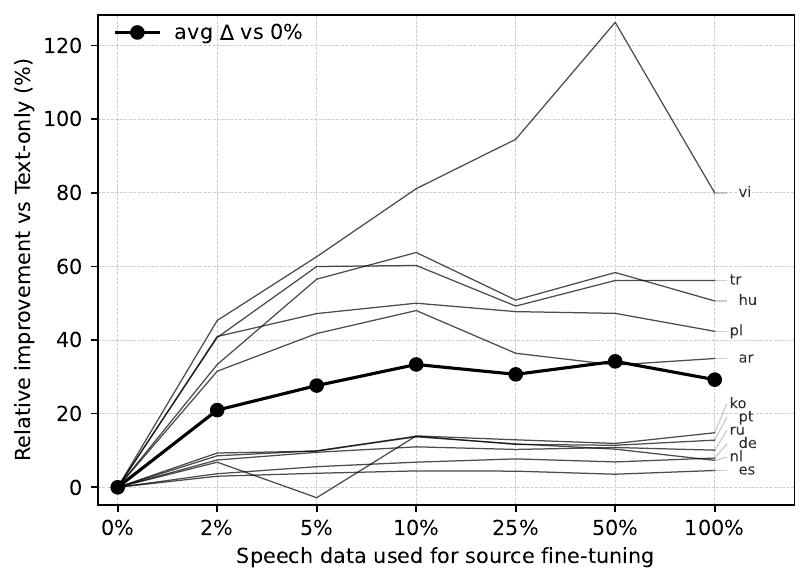}
    \caption{Zero-shot cross-lingual SLU from French to eleven target languages, reported as relative improvement in SLU-F1 over the text-only scheme.}
    \label{fig:zeroshot}
    \vspace{-0.5em}
\end{figure}

\begin{table}[t]
\centering
\caption{Few-shot SLU-F1 results using Speech-MASSIVE FR as the source and evaluation on 5 target languages.
Target fine-tuning is shown for Transcripts-only (T) and Transcripts+Speech (T+S).
Rows with +M indicate use of target MASSIVE data.}
\label{tab:fewshot}
\resizebox{\linewidth}{!}{%
\begin{tabular}{c c c c c c c}
\toprule
Source Speech & Target
& \textbf{de}
& \textbf{es}
& \textbf{ko}
& \textbf{pt}
& \textbf{vi} \\
\midrule
No Source & T   & 0.2246 & 0.2183 & 0.1084 & 0.1712 & 0.0173 \\
0\%       & T   & 0.6145 & 0.6282 & 0.4319 & 0.5418 & 0.0839 \\
2\%       & T   & 0.6378 & 0.6499 & 0.4450 & 0.5790 & 0.1158 \\
5\%       & T   & 0.6515 & 0.6548 & 0.4522 & 0.5835 & 0.1213 \\
10\%      & T   & 0.6476 & 0.6561 & 0.4609 & 0.5972 & 0.1333 \\
25\%      & T   & 0.6691 & 0.6615 & 0.4612 & 0.6085 & 0.1398 \\
50\%      & T   & 0.6645 & 0.6552 & 0.4564 & 0.5866 & 0.1529 \\
100\%     & T   & 0.6739 & 0.6683 & 0.4726 & 0.6155 & 0.1556 \\
\midrule
No Source & T+S & 0.3396 & 0.3240 & 0.2089 & 0.2842 & 0.0486 \\
0\%       & T+S & 0.6372 & 0.6630 & 0.4939 & 0.6349 & 0.3317 \\
2\%       & T+S & 0.6537 & 0.6502 & 0.4996 & 0.6342 & 0.2890 \\
5\%       & T+S & 0.6651 & 0.6685 & 0.4918 & 0.6412 & 0.3287 \\
10\%      & T+S & 0.6601 & 0.6696 & 0.4966 & 0.6436 & 0.3265 \\
25\%      & T+S & 0.6785 & 0.6721 & 0.5119 & 0.6501 & 0.3321 \\
50\%      & T+S & 0.6727 & 0.6726 & 0.5126 & 0.6588 & 0.3269 \\
100\%     & T+S & 0.6787 & 0.6804 & 0.5138 & 0.6503 & 0.3351 \\
\midrule
\multirow{2}{*}{No Source} & T+M   & 0.6551 & 0.6666 & 0.4489 & 0.5673 & 0.0692 \\
          & T+S+M & 0.6883 & 0.6708 & 0.5403 & 0.6642 & 0.3208 \\
\multirow{2}{*}{0\%}       & T+M   & 0.6600 & 0.6717 & 0.4534 & 0.5737 & 0.0730 \\
          & T+S+M & 0.6967 & 0.6835 & 0.5448 & 0.6669 & 0.3521 \\
\multirow{2}{*}{100\%}    & T+M   & 0.7057 & 0.7100 & 0.4997 & 0.6377 & 0.1718 \\
          & T+S+M & 0.7319 & 0.7226 & 0.5665 & 0.6946 & 0.3737 \\
\bottomrule
\end{tabular}
}%
\end{table}

\subsection{Results in Cross-lingual SLU}
\label{ssec:results_cross}

We now extend our analysis to the \emph{cross-lingual SLU} setting. Figure~\ref{fig:zeroshot} shows zero-shot cross-lingual evaluation from French to eleven unseen languages, reported as relative improvement over the text-only baseline. Adding 2–10\% of source-language speech data already yields gains of 20–33\% on average, with improvements saturating beyond 25\%. Despite variation across target languages, these results show that increasing source-language speech data strengthens cross-lingual grounding and directly improves generalization to other languages, even without target-language data.

We next consider the few-shot setting, focusing on five target languages (German, Spanish, Korean, Portuguese, Vietnamese), selected to span similarity to French and resource availability.
Training combines source- and target-language data: when only text data are available, we adopt a text-only scheme, while the presence of speech data leads to curriculum learning. 

Table~\ref{tab:fewshot} summarizes the results. Overall, we observe consistent improvements with more source-language speech data, with gains persisting across the full range. 
This indicates a practical path for low-resource languages: under limited target-language data, combining abundant source-language speech data can substantially improve SLU performance. 
Vietnamese is a notable exception, likely due to its larger typological gap from French and its limited representation in the model’s pretraining corpus.
In the target-language, transcripts alone provide a reasonable baseline, but adding even small amounts of speech data yields further improvements when combined.

We additionally use the MASSIVE corpus~\cite{fitzgerald2023massive}, which provides abundant target-language NLU data without speech (11,514 utterances per language), to test whether extra text-only data can further aid training.
Regardless of whether the source-language setup used transcripts alone or both transcripts and speech, adding target-language MASSIVE text yields clear improvements.
Models trained solely on MASSIVE text and Speech-MASSIVE data, without any source-language data, achieve reasonable results but still lag behind setups that leverage source-language data, underscoring its cross-lingual value.

\section{Conclusion}
\label{sec:typestyle}

In this work, we analyzed fine-tuning schemes for LALMs using SLU as a benchmark. Our monolingual experiments show that transcription-based training combined with small amounts of speech data provides substantial gains, and that curriculum learning is particularly effective in low-resource settings. In cross-lingual scenarios, source-language speech data plays a crucial grounding role, and combining it with target-language text and minimal target-language speech data further improves adaptation in low-resource languages. These findings offer practical takeaways for fine-tuning LALMs, such as prioritizing small amounts of paired speech–label data and leveraging abundant text resources. Nonetheless, our experiments are limited to a single model, Qwen2-Audio-7B-Instruct, and future work is needed to verify generalization across other LALMs and downstream tasks.

\vfill\pagebreak

\bibliographystyle{IEEEtran}
\bibliography{strings,refs}

\end{document}